# Degradation of β-$Ga_2O_3$ Schottky barrier diode under swift heavy ion irradiation*

Wen-Si Ai (艾文思)[1,2], Jie Liu (刘杰)[1,2†], Qian Feng (冯倩)[3‡], Peng-Fei Zhai (翟鹏飞)[1,2], Pei-Pei Hu (胡培培)[1,2], Jian Zeng (曾健)[1,2], Sheng-Xia Zhang (张胜霞)[1,2], Zong-Zhen Li (李宗臻)[1,2], Li Liu (刘丽)[1,2], Xiao-Yu Yan (闫晓宇)[1,2], and You-Mei Sun (孙友梅)[1,2]

[1]*Institute of Modern Physics, Chinese Academy of Sciences (CAS), Lanzhou 730000, P.R. China.*
[2]*School of Nuclear Science and Technology, University of Chinese Academy of Sciences, Beijing 100049, P.R. China*
[3]*State Key Discipline Laboratory of Wide Band Gap Semiconductor Technology, School of Microelectronics, Xidian University, Xi'an 710071, P.R. China.*

The electrical characteristics and microstructures of β-$Ga_2O_3$ Schottky barrier diode (SBD) devices irradiated with swift heavy ions (2096 MeV Ta ions) have been studied. It was found that β-$Ga_2O_3$ SBD devices showed the reliability degradation after irradiation, including turn-on voltage $V_{on}$, on-resistance $R_{on}$, ideality factor n and the reverse leakage current density $J_r$. In addition, the carrier concentration of the drift layer was decreased significantly and the calculated carrier removal rates were $5\times10^6$ - $1.3\times10^7$ $cm^{-1}$. Latent tracks induced by swift heavy ions were observed visually in the whole β-$Ga_2O_3$ matrix. Furthermore, crystal structure of tracks was amorphized completely. The latent tracks induced by Ta ions bombardments were found to be the reason for the decrease in carrier mobility and carrier concentration. Eventually, these defects caused the degradation of electrical characteristics of the devices. By comparing the carrier removal rates, the β-$Ga_2O_3$ SBD devices were more sensitive to swift heavy ions irradiation than SiC and GaN devices.

* Project supported by the National Natural Science Foundation of China (Grant Nos. 12035019, 11690041 and 12075290), China National Postdoctoral Program for Innovative Talents (Grant No. BX20200340), China Postdoctoral Science Foundation (Grant No. 2020M673539), CAS "Light of West China" Program and the Youth Innovation Promotion Association of CAS (Grant No. 2020412)

† Corresponding author. E-mail address: j.liu@impcas.ac.cn (Jie Liu)
‡ Corresponding author. E-mail address: qfeng@mail.xidian.edu.cn (Qian Feng)

## 1. Introduction

Monoclinic $Ga_2O_3$ ($\beta$-$Ga_2O_3$) is a traditional transparent conductive oxide materials and $\beta$-$Ga_2O_3$ based photodetectors are attracting interest as truly solar-blind deep ultraviolet photodetectors, since they exhibit cut-off wavelengths below 280 nm.[1,2] This makes $\beta$-$Ga_2O_3$ attractive in the fields of new generation photoconductors, such as deep ultraviolet detectors, light-emitting diodes and lasers. The research on $\beta$-$Ga_2O_3$ has been extremely hot in the past decade because of its new application in power electronic devices. $\beta$-$Ga_2O_3$ has not only excellent optical properties, but also a large bandgap of 4.7-4.9 eV and a high critical electric field strength of 8 MV/cm.[3] The large bandgap and the high critical electric field strength enables $\beta$-$Ga_2O_3$ based devices to operate at high temperature and high power. Furthermore, $\beta$-$Ga_2O_3$ can be prepared by melting method, which was the same as Si and sapphire substrate. Compared with SiC and GaN, the cost advantage of $\beta$-$Ga_2O_3$ further promotes its application in the field of high-power electronic devices.[4]

$\beta$-$Ga_2O_3$ devices will face huge challenges used in aerospace systems despite their excellent properties. The radiation environment in outer space comprises high-energy protons, electrons, neutrons, and heavy ions.[5] Then, the different types of damages can be formed in the devices after different particle irradiation. For electrons, protons and $\gamma$-rays irradiation, simple point defects are generally introduced in the wide band gap semiconductors.[6,7] Heavy ions and fast neutrons mainly introduce point defects or cascade displacement damages by elastic collision with target atoms.[8] $\beta$-$Ga_2O_3$ is generally considered to be radiation hardness to displacement damage due to the high bond energy and large band gap.[9] According to the literature, the 4H-SiC single crystal was amorphous at fluence of 0.4 dpa (displacements per atom) for 4 MeV Xe ions irradiation,[10] but the saturate disorder state of $\beta$-$Ga_2O_3$ single crystal can be reached at a higher fluence of 0.6 dpa for 700 keV Sn ions irradiation.[11] Moreover, the irradiation response of carrier concentration in $\beta$-$Ga_2O_3$ Schotty barrier diode (SBD) is similar to that of GaN devices after irradiated by electrons and protons.[12]

Different from the above traditional particles that mainly introduce damages by interaction with the target atoms, the swift heavy ions (SHIs, >1 MeV/u), one of the cosmic rays, mainly transfer energy to the target electrons through huge electronic energy deposition and target electrons further transfer the energy to the atoms through electron-phonon coupling.[13,14] When electronic energy loss ($S_e$) is lager enough, a single swift heavy ion can cause local melting of material and introduce amorphous or recrystallized damage region during quenching. This damage region of nanometer in size is called latent track. In our previous study, it was found that amorphous latent tracks can be introduced in β-$Ga_2O_3$ single crystal when the $S_e$ exceeded 17 keV/nm.[15] However, the effect of latent tracks on the electrical characteristics of β-$Ga_2O_3$ devices is still not yet studied. Therefore, 2096 MeV Ta ions were used to irradiate β-$Ga_2O_3$ SBD devices in this work and the role of latent tracks on the reliability degradation of devices was analyzed in detail.

## 2. Experimental details

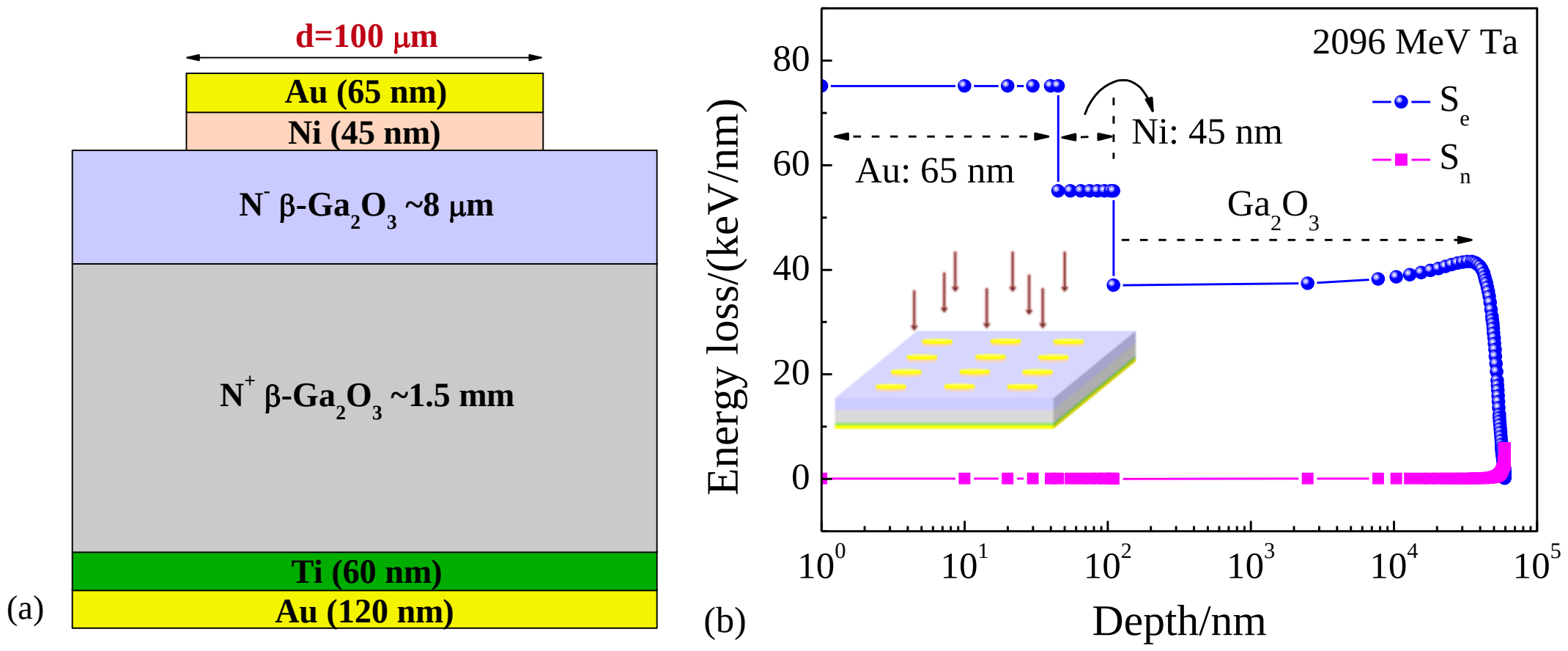

Fig.1. (color online) (a) The schematic across section of the vertical β-$Ga_2O_3$ SBD and (b) the distribution diagram of electronic energy loss ($S_e$) and nuclear energy loss ($S_n$) of 2096 MeV Ta in β-$Ga_2O_3$ SBD.

The vertical β-$Ga_2O_3$ SBD devices were used in this work. The $N^-$ β-$Ga_2O_3$ (001) drift layer (Sn: ~$1.8\times10^{16}$ $cm^{-3}$) of thickness 8 μm was deposited by hydride vapor phase epitaxy on 1.5 mm bulk $N^+$ substrate (Sn: ~$3\times10^{18}$ $cm^{-3}$). The metal stack of Ti/Au was deposited on the whole back of $N^+$ substrate by E-beam evaporation and followed by the rapid thermal annealing at 500℃ for 60 s under nitrogen atmosphere to form the Ohmic contact. The front side of $N^-$ drift layer was patterned by lift-off of E-beam deposited Schottky contacts Ni/Au (45 nm/65 nm). The diameter of Schottky contact was about 100 μm. The structure of the schematic across section of β-$Ga_2O_3$ SBD was shown in Fig. 1 (a). The β-$Ga_2O_3$ SBD devices were divided into three groups and named #1, #2 and #3, respectively.

Heavy ion irradiation experiment was performed at the Heavy Ion Research Facility in Lanzhou (HIRFL) in the Institute of Modern Physics (IMP), Chinese Academy of Sciences (CAS). The β-$Ga_2O_3$ SBD devices without electrical bias were irradiated with 2096 MeV Ta ions in the vacuum chamber. The $S_e$ and nuclear energy loss ($S_n$) were calculated by SRIM 2013 code[16] and the detail distribution of $S_e$ and $S_n$ in β-$Ga_2O_3$ SBD was plotted in Fig. 1 (b). The range of 2096 MeV Ta ions in the device is about 50 μm, reaching deep inside the substrate far away from the Metal-Semiconductor (M-S) interface. Due to the limited number of samples, cumulative irradiation was adopted in this work. The devices were irradiated for the first time with the fluence of $5\times10^7$-$5\times10^8$ ions/cm$^2$, respectively. After irradiation, the irradiated samples were removed from the vacuum chamber for electrical properties measurement. Then they were continued to irradiated until the fluence up to $1\times10^9$-$1\times10^{10}$ ions/cm$^2$, respectively. The specific irradiation parameters were listed in Table 1.

Table 1. The irradiation fluence of the three groups of β-$Ga_2O_3$ SBD in the first irradiation experiment and the total fluence after the second cumulative irradiation experiment.

| Irradiation batches | #1 | #2 | #3 |
|---|---|---|---|
| 1$^{st}$ | $5\times10^7$ ions/cm$^2$ | $1\times10^8$ ions/cm$^2$ | $5\times10^8$ ions/cm$^2$ |
| 2$^{nd}$ | $1\times10^{10}$ ions/cm$^2$ | $5\times10^9$ ions/cm$^2$ | $1\times10^9$ ions/cm$^2$ |

Current density-voltage (J-V) and high frequency (1 MHz) capacitance-voltage (C-V) characteristics were measured by a Keithley 4200 Semiconductor Parameter

Analyzer at room temperature. For each fluence, ten Schottky electrodes at least with almost identical electrical characteristics were analyzed. The normal behaviors of J-V and C-V are shown in the next section. The microstructure of β-$Ga_2O_3$ SBD after irradiation was characterized by bright-field TEM using a Tecnai G2 F20 S-TWIN TEM (FEI, USA) at the accelerating voltage of 200 kV.

## 3. Results and Discussion

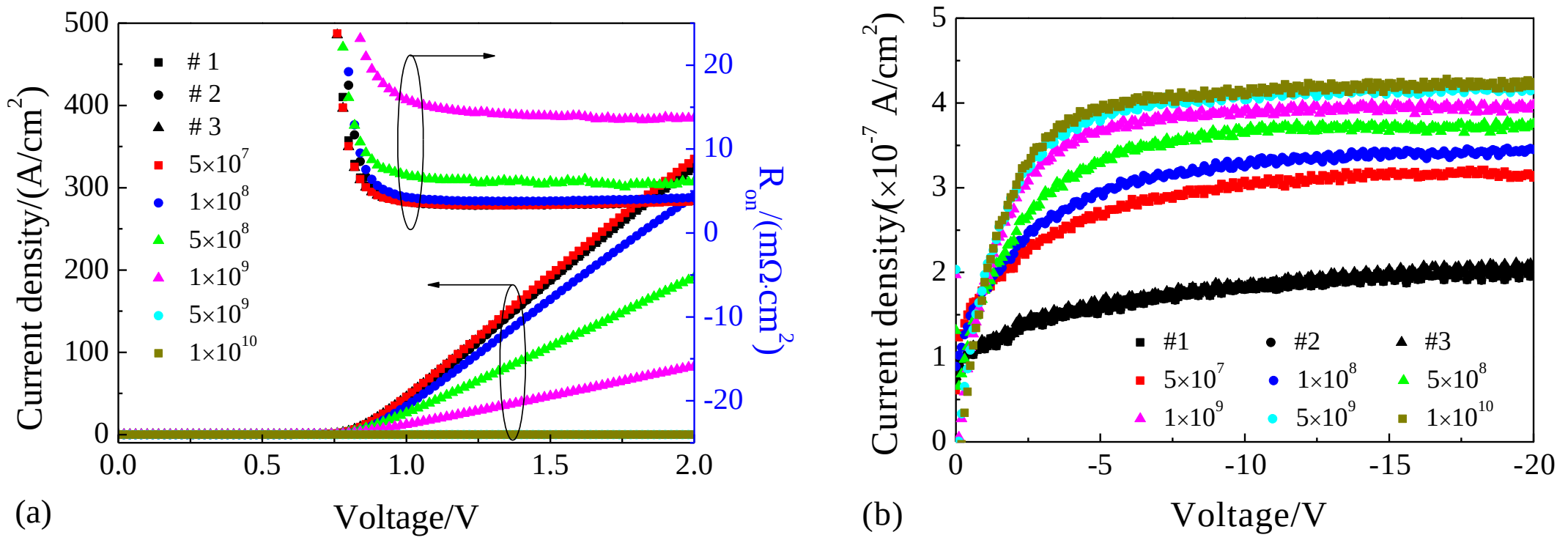

Fig.2. (color online) (a) The forward J-V characteristics and the differential on-resistance $R_{on}$ before and after irradiation. (b) The reverse J-V characteristics before and after irradiation. The unit of Ta ions fluence is ions/cm$^2$.

Fig. 2 (a) shows the forward J-V characteristics and the differential on-resistance $R_{on}$ as a function of the voltage for β-$Ga_2O_3$ SBD devices with different ion fluences. The results show that the forward current density decreases gradually with the increase in fluence. At the forward bias of 2 V, the maximum current density decreased from 327 A/cm$^2$ to 83 A/cm$^2$ and the $R_{on}$ increased from 3.8 mΩ·cm$^2$ to 13.7 mΩ·cm$^2$ at the fluence of $1\times10^9$ ions/cm$^2$. When the ion fluence increased to $5\times10^9$ and $1\times10^{10}$ ions/cm$^2$, the β-$Ga_2O_3$ SBD devices do not exhibit forward guide characteristics and the $R_{on}$ values reach to the order of MΩ·cm$^2$ (see Table 2). The reverse J-V characteristic also indicates the increase of reverse leakage current density as shown in Fig. 2 (b). It suggests that Ta ions irradiation can significantly affect the J-V characteristics of the β-$Ga_2O_3$ SBD devices and degrade the performance.

According to the thermionic emission theory,[17,18] the relationship between the voltage and the current density can be described as:

$$J = J_s exp\left(\frac{qV}{nkT}\right), \tag{1}$$

$$J_s = A^* T^2 exp\left(-\frac{\Phi_B}{kT}\right), \tag{2}$$

where $J_s$ is the saturation current density, n is the ideality factor, k is Boltzmann's constant, T is the absolute temperature, $A^*$ is the effective Richardson constant (41.1 $A/(cm^2 K^2)$), and $\Phi_B$ is the Schottky barrier height. The parameters n and $\Phi_B$ can be estimated by fitting the linear region of the J-V curve and the detail electrical parameters of $\beta$-$Ga_2O_3$ SBD devices before and after irradiation were summarized in Table 2. In order to compare the variation of electrical parameters more intuitively, the increment of each parameter (the parameter value after irradiation minus the parameter value before irradiation) is shown in Fig. 3. Since the $\beta$-$Ga_2O_3$ SBD devices do not exhibit forward guide characteristics when the fluence is up to $5\times10^9$ and $1\times10^{10}$ ions/cm$^2$, the variation of $V_{on}$, n and $\Phi_B$ in Fig. 3 (a) only covered in the fluence range from $5\times10^7$ to $1\times10^9$ ions/cm$^2$.

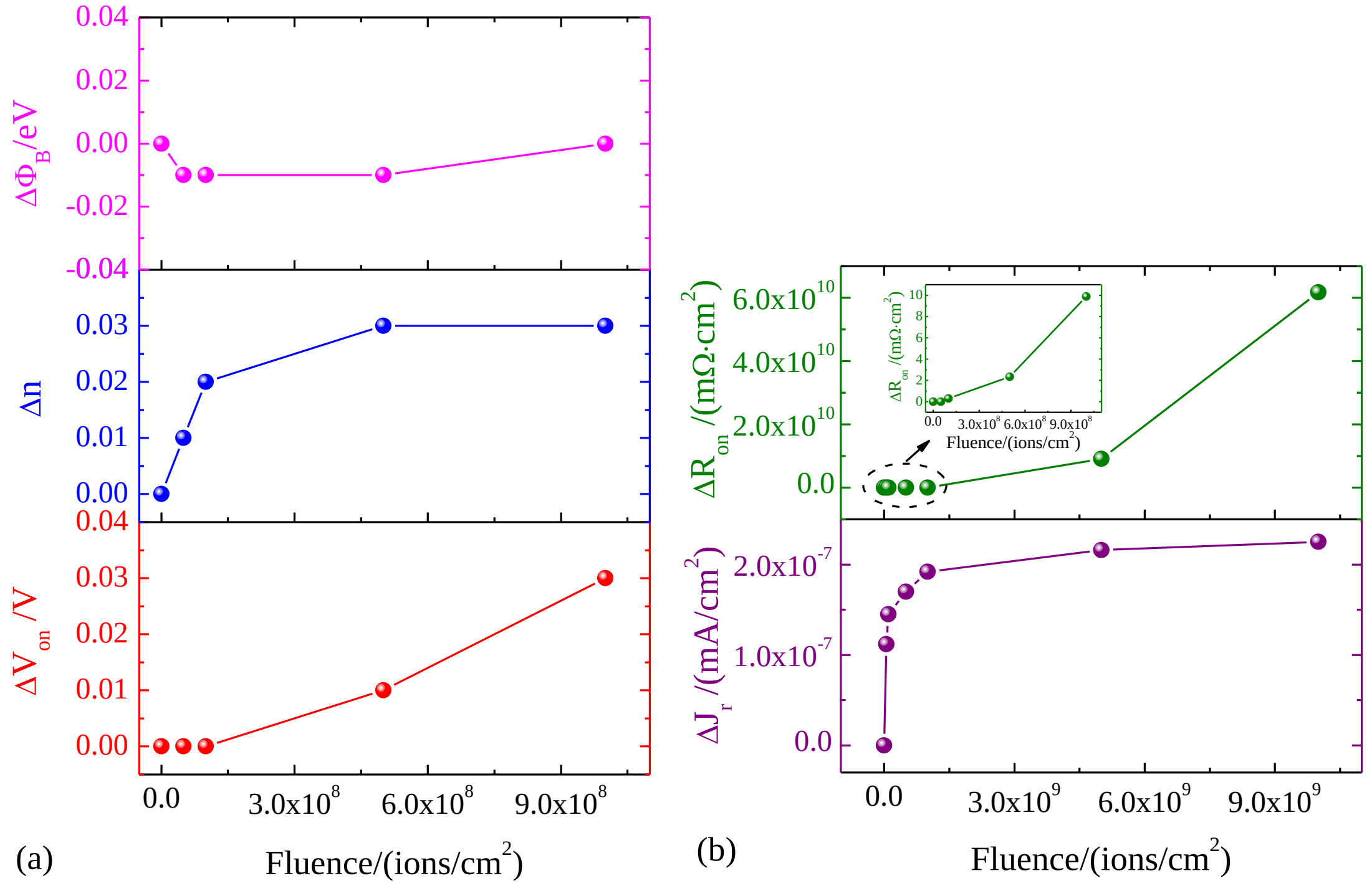

Fig.3. (color online) The increments of electrical parameters (turn-on voltage $V_{on}$, ideality factor n, Schottky barrier height $\Phi_B$, reverse leakage current density $J_r$ and on-resistance $R_{on}$) as a function of fluence before and after irradiation.

Table 2. The comparison of experimentally calculated values of β-$Ga_2O_3$ SBD devices before and after 2096 MeV Ta ions irradiation

| Parameters | Pre-irradiation | | | $5\times10^7$ ions/cm$^2$ | $1\times10^8$ ions/cm$^2$ | $5\times10^8$ ions/cm$^2$ | $1\times10^9$ ions/cm$^2$ | $5\times10^9$ ions/cm$^2$ | $1\times10^{10}$ ions/cm$^2$ |
|---|---|---|---|---|---|---|---|---|---|
| | #1 | #2 | #3 | | | | | | |
| $V_{on}$ (V) | 0.75 | 0.78 | 0.75 | 0.75 | 0.78 | 0.76 | 0.78 | —— | —— |
| $R_{on}$ at 2 V (mΩ cm$^2$) | 3.83 | 3.88 | 3.82 | 3.83 | 4.18 | 6.16 | 13.71 | $9.07\times10^9$ | $6.17\times10^{10}$ |
| n | 1.01 | 1.02 | 1.02 | 1.02 | 1.04 | 1.05 | 1.05 | —— | —— |
| $\Phi_B$ (eV) | 1.13 | 1.15 | 1.12 | 1.12 | 1.14 | 1.11 | 1.12 | —— | —— |
| $J_r$ at -20 V ($\times10^{-7}$ A/cm$^2$) | 2.00 | 2.00 | 2.05 | 3.12 | 3.45 | 3.75 | 3.97 | 4.16 | 4.25 |
| $N_d$-$N_a$ ($\times10^{16}$ cm$^{-3}$) | 1.81 | 1.79 | 1.81 | 1.81 | 1.74 | 1.15 | 0.54 | —— | —— |
| $R_c$ (cm$^{-1}$) | —— | —— | —— | 0 | $5\times10^6$ | $1.3\times10^7$ | $1.3\times10^7$ | —— | —— |

In Fig. 3 (a), both of the turn-on voltage $V_{on}$ and ideality factor n increased with fluence increasing, while the parameter $\Phi_B$ showed little changes. The increase of ideality factor n indicates that the current transport mechanism gradually deviates from the thermionic emission model. In general, the defects introduced by irradiation can lead to the increase of M-S interface state density and then other current transport mechanisms will participate in the process, such as tunneling.[19,20] The defects can also act as the capture centers of carriers, resulting in the decrease of the carrier concentration and mobility.[21] Hence, the on-resistance $R_{on}$ value increased with the increase in fluence as shown in Fig. 3 (b). In general, the reverse leakage current density $J_r$ can reflect the blocking characteristic of SBD. In Fig. 3 (b), the increase of $J_r$ after irradiation indicates the degradation of blocking. This is mainly related to the reduced of carrier lifetime due to the increase of deep level recombination centers in the barrier region after irradiation.[22]

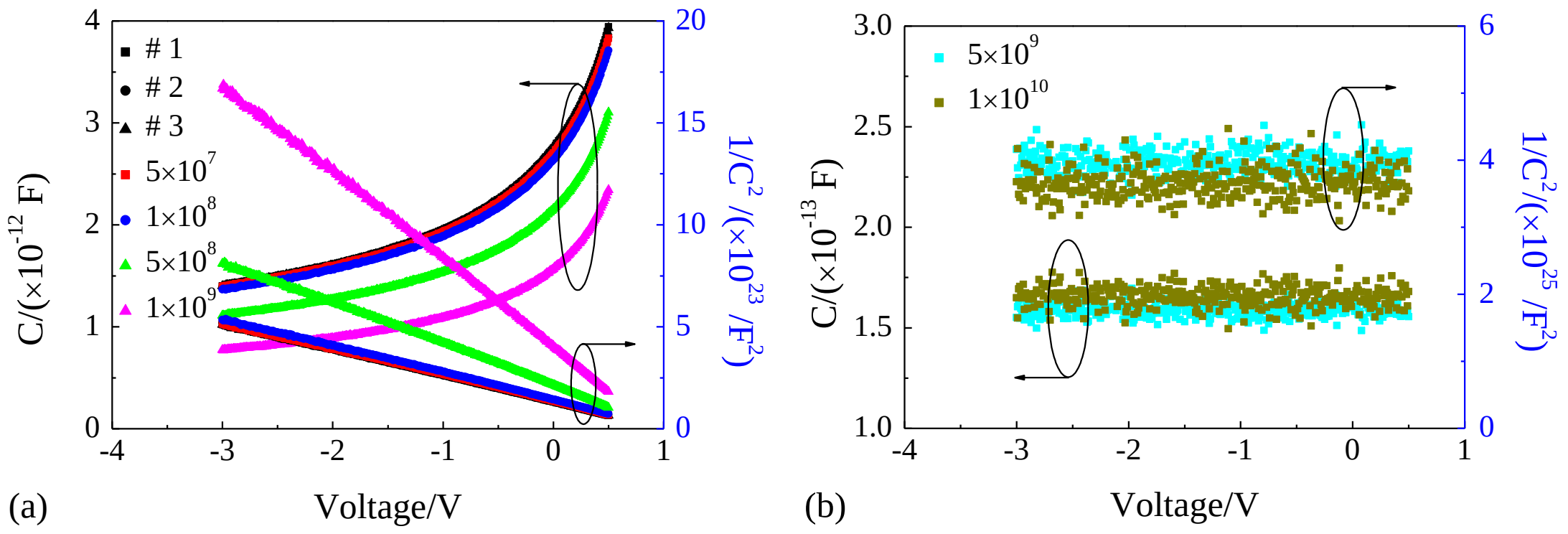

Fig.4. (color online) C-V and $1/C^2$-V characteristics (1 MHz) of the devices after 2096 MeV Ta ions irradiation. The unit of Ta ions fluence is ions/cm$^2$.

Fig. 4 shows the C-V and $1/C^2$-V plots at a frequency of 1 MHz. The C-V relationship for a Schottky barrier is:[23]

$$\frac{1}{C^2} = \frac{2}{q\varepsilon A^2(N_d - N_a)}(V_{bi} - V), \qquad (3)$$

where q is the electron charge, A is the area of the Schottky diode, ε is the dielectric constant ($Ga_2O_3$, $\varepsilon=10\varepsilon_0$), $V_{bi}$ is the built-in potential, ($N_d$-$N_a$) stands for the carrier concentration in the drift layer. The carrier concentration can be extracted from the slope of the $1/C^2$-V curve and the results were listed in Table 2. Only the carrier concentrations in the drift layer were calculated with the fluence range of $5\times10^{7}$-$1\times10^{9}$ ions/cm$^2$. The corresponding variation of normalized carrier concentration was summarized in Fig. 5 (a).

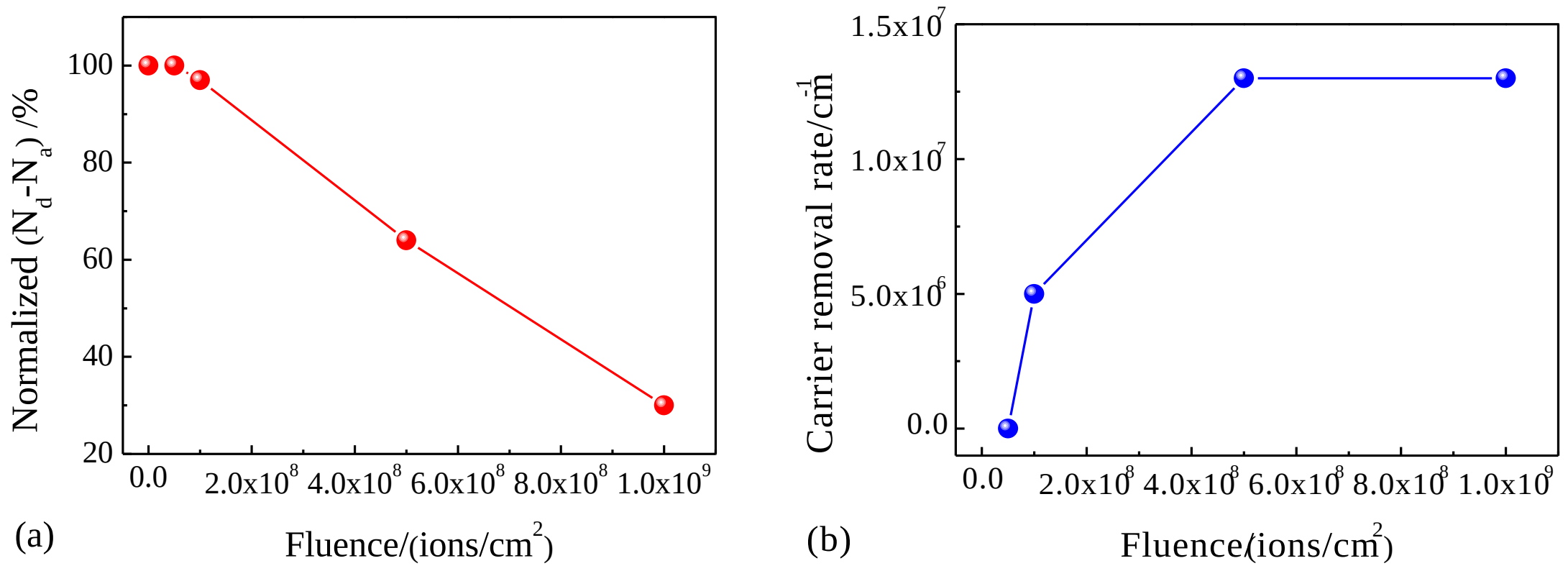

Fig.5. (color online) (a) The normalized carrier concentration and (b) carrier removal rate in the drift β-$Ga_2O_3$ layer after 2096 MeV Ta ions irradiation.

According to Fig. 5 (a), it is clear that the carrier concentration in the drift β-$Ga_2O_3$ layer shows little changes at the fluence of $5\times10^{7}$ ions/cm$^2$. However, the carrier concentration decreased significantly with the fluence increased from $1\times10^{8}$ to $1\times10^{9}$ ions/cm$^2$. At the fluence of $1\times10^{9}$ ions/cm$^2$, the normalized carrier concentration is only 30% that of the unirradiated samples. The acceptor-defects introduced by Ta ions result in the decrease of carrier concentration, further cause the increase of depletion width, and finally show that the capacitance in C-V measurement decreases with the increase of fluence. As the fluence increases further to $5\times10^{9}$ and $1\times10^{10}$ ions/cm$^2$, the excessively low carrier concentration is equivalent to the infinite width of the depletion layer and the Schottky barrier capacitance disappeared.

According to the carrier concentration, the carrier removal rate $R_c$ was calculated and the results were plotted as shown in Fig. 5 (b). The carrier removal rate $R_c$ relates to the removal of carriers as deep traps which were introduced by radiation. It is related to the fluence $\varphi$ and the decrease value of carrier concentration $\Delta(N_d-N_a)$, through the equation:[6,24]

$$R_c\varphi = \Delta(N_d - N_a). \tag{4}$$

The $R_c$ can provide a practical guide for estimating the degree of the degradation induced in the devices or materials for a given fluence of the common type of radiation. In this work, the calculated $R_c$ was $5\times10^6$ $cm^{-1}$ for β-$Ga_2O_3$ SBD irradiated with Ta ions to the fluence of $1\times10^8$ ions/$cm^2$ and it reached saturation values of $1.3\times10^7$ $cm^{-1}$ at the fluence of $5\times10^8$ ions/$cm^2$. In general, $R_c$ is linear increasing with the fluence at the lower fluence. However, if most of the carriers are removed at a higher fluence, the excess defects will not contribute to the carrier removal effect any more. Thus, the relationship between $R_c$ and the fluence φ will not follow the linear relationship.[6]

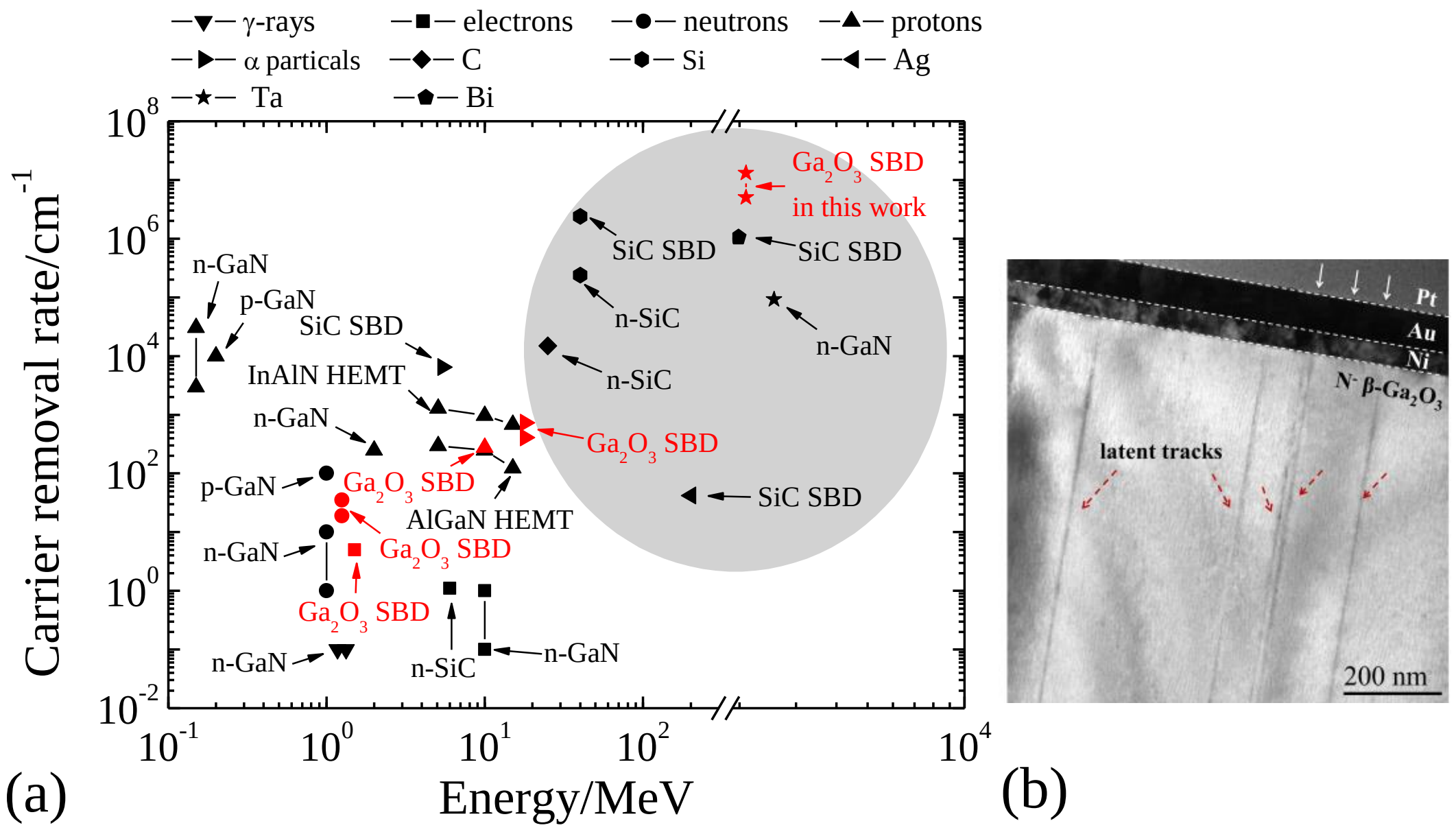

Fig.6. (color online) (a) Carrier removal rate summary diagram in β-$Ga_2O_3$ (red symbols)[12,25–27] and other types of GaN or SiC based devices (black symbols)[6,22,28–32] with different species and energies ion-irradiation. The shadow represents the energy regions of swift heavy ions. The data of red star are from our work. (b) The cross-sectional TEM image of β-$Ga_2O_3$ SBD irradiated with 2096 MeV Ta ions to a fluence of $1\times10^{10}$ ions/$cm^2$. The irradiation direction is indicated by white arrows and the latent tracks parallel to each other are marked by red arrows.

The carrier removal rates of β-$Ga_2O_3$ based devices irradiated by different types of ions were summarized as shown in Fig. 6 (a) (red symbols).[12,25–27] Note that carrier removal rate is 406-728 $cm^{-1}$ for α particles irradiation,[27] 300 $cm^{-1}$ for 10 MeV protons,[12] 4.9 $cm^{-1}$ for 1.5 MeV electrons[26] and 19-28 $cm^{-1}$ for 1.25 MeV neutrons[25] in β-$Ga_2O_3$ SBD devices or rectifiers. However, the carrier removal rates for SHIs in this work are much higher. This indicates that the energetic Ta ions exhibit the highest carrier removal rates among these ions irradiation, and it can be explained by the damage type caused by SHIs.

The cross-sectional TEM of the β-$Ga_2O_3$ SBD irradiated by 2096 MeV Ta ions to a fluence of $1\times10^{10}$ ions/$cm^2$ was shown in Fig. 6 (b). It can be seen that the interface between Ni and $N^-$ β-$Ga_2O_3$ layer is sharp and there is little inter-mixting at the highest fluence of $1\times10^{10}$ ions/$cm^2$. However, we found indication of latent tracks parallel to each other at $N^-$ β-$Ga_2O_3$ layer. In our previous work,[15] TEM results proved that one single 2096 MeV Ta ion introduced the amorphous latent track with a size of ~8 nm in β-$Ga_2O_3$ single crystal. Considering the range of 2096 MeV Ta ions in β-$Ga_2O_3$ SBD devices, the latent tracks can be introduced not only at the 8 μm $N^-$ layer, but also within the range of 40 μm at the $N^+$ layer. For a single swift heavy ion irradiation, the latent track along the ion trajectory is a nanometer-size amorphous region. For a single proton or α particle irradiation, the introduced damage is isolated atomic-size point defects. Hence Ta ions exhibit the highest carrier removal rate.

Fig. 6 (a) also summarized the carrier removal rates of GaN or SiC based devices including SBD devices and high electron mobility transistors (HEMTs).[6,22,28–32] It can be extracted from Fig. 6 (a) that under the irradiation environment of high-energy electrons, protons and heavy ions which are mainly introduced displacement damages by elastic collision with the target atoms, the $R_c$ values of β-$Ga_2O_3$ SBD or rectifier are similar to that of GaN or SiC based devices, indicating the excellent radiation hardness of β-$Ga_2O_3$ devices. This can be attributed to the higher formation energy of vacancy defects in β-$Ga_2O_3$.[33–35] However, the degradation of β-$Ga_2O_3$ SBD is more serious than that of SiC or GaN devices under the SHIs irradiation as the shadow shown in Fig. 6 (a). In addition, $Ga_2O_3$ SBD devices in our work are completely damaged under 2096 MeV Ta ions irradiation with fluence of $5\times10^9$ ions/cm$^2$. However, the GaN HEMTs reported by Hu et al.[36] are still functional after swift heavy Bi ions irradiation with energy of 1500 MeV to the fluece of $1.7\times10^{11}$ ions/cm$^2$.

Based on the thermal spike model,[37] the latent track is formed through the material melting and quenching rapidly along the path of SHIs. Hence, thermodynamic properties and recrystallization ability of the target material are the main factors affecting the latent track formation.[38] The poor thermal conductivity and recrystallization ability of β-$Ga_2O_3$ make the $S_e$ threshold of latent track formation in β-$Ga_2O_3$ (17 keV/nm) is lower than that of SiC (>34 keV/nm) and GaN (23-28 keV/nm).[15] Therefore, the damage introduced by SHIs in the whole β-$Ga_2O_3$ matrix has a greater impact on the degradation of β-$Ga_2O_3$ SBD devices than the damage in M-S interface.

## 4. Conclusions

In conclusion, we studied the degradation and the structure damages of β-$Ga_2O_3$ SBD devices after 2096 MeV Ta ions irradiation with the fluence range from $5\times10^7$ to $1\times10^{10}$ ions/cm$^2$. Both the conducting and blocking characteristics were sensitive to ion irradiation. A strong reduction of the carrier was observed and the carrier removal rates were $5\times10^6$ -$1.3\times10^7$ cm$^{-1}$. Furthermore, the amorphous latent tracks along the ions trajectories cross the whole area of drift layer, were responsible for the decrease in carrier concentration and mobility, and resulted in the deterioration of the β-$Ga_2O_3$ SBD devices. In addition, the damage introduced by SHIs in the whole β-$Ga_2O_3$ matrix has a greater impact on the degradation of β-$Ga_2O_3$ SBD devices than the damage in M-S interface. The serious degradation for β-$Ga_2O_3$ SBD indicates the worse radiation hardness of β-$Ga_2O_3$ based device to SHIs compared with SiC and GaN devices.

## Acknowledgments

The authors would like to thank the accelerator staff of HIRFL of IMP for the technical support and for providing the characterization facilities. We also thank Dr. Lijun Xu at the Institute of Modern Physics, Chinese Academy of Sciences for support of the TEM measurement.